\documentclass[amsmath,amssymb,nofootinbib,prb,10pt,floatfix]{revtex4}
\usepackage{bm}
\usepackage{amssymb}
\usepackage{graphicx}
\usepackage{epstopdf}

\begin{document}

\title{Multipole structure and coordinate systems}

\author{Lior M.~Burko \\ Department of Physics \\
University of Alabama in Huntsville, Huntsville, Alabama 35899}

\date{January 2, 2007}


\begin{abstract} 
Multipole expansions depend on the coordinate system, so that coefficients of multipole moments can be set equal to zero by an appropriate choice of coordinates. Therefore, it is meaningless to say that a physical system has a nonvanishing quadrupole moment, say, without specifying which coordinate system is used. (Except if this moment is the lowest non-vanishing one.) 
This result is demonstrated for the case of two equal like electric charges. Specifically, an adapted coordinate system in which the potential is given by a monopole term only is explicitly found, the coefficients of all higher multipoles vanish identically.  It is suggested that this result can be generalized to other potential problems, by making equal coordinate surfaces coincide with the potential problem's equipotential surfaces. 
\end{abstract}


\maketitle

\section{Introduction}
Multipole expansions are a standard mathematical tool in both physics research and  teaching (e.g., Ref.~\cite{Morse}). For example, students of electromagnetism often calculate the electrostatic potential of a point electric charge when the coordinate system is not centered on the latter \cite{Griffiths,Panofsky}. In mathematical methods courses this problem often motivates, and serves as a means to the introduction of Legendre functions through its generating function 
$$
g(t,x):=(1-2xt+t^2)^{-1/2}=\sum_{n=0}^{\infty}P_n(x)\, t^n\;\;\;\;\;\;\;,\;\;\;\;\;\;|t|<1\, .
$$
(The alternative approach to the introduction of the Legendre functions is through the angular equation obtained from the separation of variables of the Poisson equation in spherical symmetry \cite{Jackson, Pollack,Franklin,Reitz}.) 
As the generating function is equal to the inverse distance between two points (in three-dimensional Euclidean space)---i.e., $1/|{\mathbf x}-\mathbf{x}'|=g(r'/r,\,\cos\theta)$, where $\mathbf{x}$,$\mathbf{x}'$ are the two points---one can readily superpose solutions to obtain the field of an arbitrary distribution of point charges. Here, $r,r'$ are their corresponding distances from the origin (so that $r'<r$) and $\theta$ is the angle between the vectors $\mathbf{x}$ and $\mathbf{x}'$ \cite{Mathews}.   
The electrostatic potential then is expressed as an infinite sum over Legendre functions (see, e.g., 
\cite{Jackson}), with all multipoles present (both even and odd). 

The next problem in a typical mathematical methods course may often be to calculate the potential of a pair of charges, either of opposite signs (``electric dipole") or of the same sign. In either case, half of the coefficients in the series vanish, and one ends up with a series over all odd or even terms, respectively. Specifically, in the latter case (two equal charges $q$), the potential is 
\begin{equation}\label{multipoles}
\phi(r,\theta)=\frac{2q}{r_>}\left[1+P_2(\cos\theta)\left(\frac{r_<}{r_>}\right)^2+P_4(\cos\theta)\left(\frac{r_<}{r_>}\right)^4+\cdots\right]\, ,
\end{equation}
where $r_>(r_<)$ is the greater (smaller) of the evaluation point and the location of the charge from the center of the coordinate system. (We assume here, without loss of generality, that the charges are on the $z$-axis, so that the potential is azimuthal.) Notably, the potential includes a monopole term, a quadrupole term, a hexadecapole term, and so on. The different multipoles are used here describe the moments of a given charge distribution. 
Specifically, the quadrupole moment equals $2qr_<^2/r_>^3$. 
When $r_>\gg r_<$, the potential approximately equals the total charge divided by the distance to the evaluation point, which is the known result for a point particle at the origin of the coordinate system. Indeed, under this condition the separation of the charge from the origin is very small (compared with the distance to the evaluation point), so that the potential is indeed expected to be just such. However, when this condition is not satisfied, or indeed at finite distances, we find that the potential needs to be corrected, the correction terms being the multipole moments. 

Mathematical methods texts often discuss next how to arrange the charge distribution so that one, or more, of the multipole moments vanishes (in a given coordinate system) 
\cite{Arfken}. As the electrostatic problem is linear, one may superpose charges so that the lowest non-vanishing multipole moment may be as desired, e.g., one may obtain an electric  dipole by having a charge $q$ at $z=a$ and a charge $-q $ at $z=-a$ thus canceling the monopole term, and the lowest non-vanishing moment is the dipole. 
An interesting point that is usually not highlighted is that the multipole  expansion itself is coordinate dependent, so that for a {\em given} charge distribution one can choose adapted coordinated in which certain multipole moments vanish.  Specifically, we will show here in detail how, for the aforementioned example of two equal, like, static electric charges, one can find an adapted coordinate system in which {\em all the multipoles above the monopole vanish}. In such a coordinate system the potential is a function of one coordinate only, as the potential includes only a monopole term, which highlights the simplicity and elegance of the presentation. We see then that the complex multipole structure of the potential (\ref{multipoles}) is a consequence of the choice of coordinates. We can find a coordinate system in which the potential equals the total charge divided by a coordinate that plays, in some loose sense, the role of ``effective distance." It is only the lowest non-vanishing moment that cannot be nullified by a coordinate transformation. 

The price to pay for this simplicity and elegance is in the form of a non-trivial coordinate transformation from, say, cartesian coordinates to the adapted ones, and in topologically non-trivial coordinates. Arguably (and only semi-seriously), there is ``conservation of complexity," which one cannot escape. However, several useful tools of mathematical physics are used in order to find this adapted coordinate system, which makes it an interesting problem, in addition to leading to deeper understanding of multipole expansions and their meaning.

\section{The Two--Distance Coordinates}

\begin{figure}
\input epsf
\includegraphics[width=12.0cm]{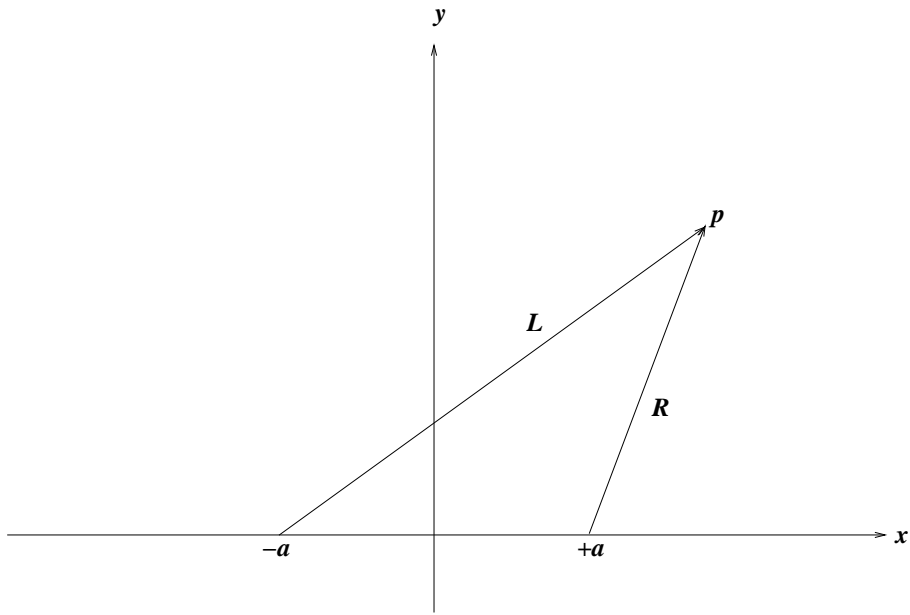}
\caption{The 3D--TDC: shown is the $x-y$ plane. The two centers are at $x=a,y=0$ and $x=-a,y=0$. The field evaluation point is at $p$. The coordinates $L$ and $R$ are defined by the distance of $p$ from the two centers. The third coordinate $\Phi$ is obtained by rotation about the $x$-axis.  }
\label{fig1}
\end{figure}

Let us first introduce a convenient coordinate system, from which we can transform to the desired adapted coordinates. Our strategy is to look for a coordinate system that is adapted to the potential in space. As the potential depends only on the distances to the two charges, a good starting point would be to base a coordinate system on the two distances from the two charges, so that equal coordinate surfaces would be spheres centered on either charge. The intersection of two such spheres spans a circle, and an azimuthal coordinate can then uniquely specify a point in space. To construct this coordinate system, position the charges along the $x$-axis, say, at $x=\pm a$. Define the distance of the evaluation point of the field from the two centers:
\begin{equation}
L:=\sqrt{(x+a)^2+y^2+z^2}\;\;\;\;\;\;R:=\sqrt{(x-a)^2+y^2+z^2}\, .
\end{equation}
These would be two coordinates, and the third one is defined by a rotation about the $x$-axis, namely, 
\begin{equation}
\Phi:=\tan^{-1}\,\left(\frac{z}{y}\right)\, .
\end{equation}
The coordinate transformation from the cartesian coordinates $x,y,z$ to the coordinates $L,R,\Phi$ is a regular coordinate transformation, as is evident from the explicit expressions for the inverse transformation: 
\begin{eqnarray}
x&=&\frac{L^2-R^2}{4a}\\
y&=&\sqrt{R^2-\frac{(L^2-R^2-4a^2)^2}{16a^2}}\,\sin\Phi :=Q\,\sin\Phi \\
z&=&\sqrt{R^2-\frac{(L^2-R^2-4a^2)^2}{16a^2}}\,\cos\Phi :=Q\,\cos\Phi
\end{eqnarray}
The coordinates $R,L,\Phi$ are the {\em Two--Distance Coordinates}, or 3D--TDC. We will find these coordinates to be a convenient starting point in our search for the adapted coordinate system, although they are unadapted to the potential problem themselves. The (spatial) metric in the 3D--TDC is given by 
\begin{equation}\label{lr-metric}
\,d\sigma^2=\frac{L^2\,R^2}{4a^2Q^2}\,\left(\,dL^2+\,dR^2\right)-\frac{LR(L^2+R^2-4a^2)}{4a^2Q^2}\,dL\,dR+Q^2\,d\Phi^2\, ,
\end{equation}
where 
\begin{equation}Q^2=R^2-\frac{(L^2-R^2-4a^2)^2}{16a^2}=\frac{[(L+R)^2-4a^2][4a^2-(L-R)^2]}{16a^2}=y^2+z^2\, .
\end{equation}
The metric (\ref{lr-metric}) is evidently non-diagonal, as it includes the ``cross term" proportional to $\,dL\,dR$. Non-diagonal metrics are of much importance in physics and mathematical physics. Nevertheless, typical mathematical method courses do not discuss them. This metric is singular at the origin (half way between the two charges)---as can be seen from the vanishing of $Q^2$ there---but as this is flat Euclidean space, this singularity is a coordinate singularity. Figure \ref{fig1} displays the 3D--TDC system. 

\section{Searching for the adapted coordinates}
To search effectively for the adapted coordinate system, let us revisit regular spherical coordinates, and see why they are so effective in the description of the electric field of a single point charge. The main feature of the electric potential of a point charge is that equipotential surfaces are concentric spheres. The latter are the loci of all equidistant points from the center. These two statements imply that  anywhere on an equipotential surface the value of one of the coordinates (the radial coordinate) is constant. To find the potential at some evaluation point, we only need to know on which equipotential surface it lies. For this reason, the potential is a function of only one coordinate, because only one coordiante is required to identify an equipotential surface. After the radial coordinate is chosen, one can readily find the other coordinates: the polar coordinate $\theta$ is found by requiring that it is everywhere orthogonal to constant-$r$ surfaces, and the azimuthal coordinate is obtained by rotation of the $r-\theta$ plane.  Spherical coordinates are therefore adapted to the potential problem of a single charge at their center, and the potential is described in full by a monopole term only. We show this derivation in detail in Appendix A.

\begin{figure}
\input epsf
\includegraphics[width=12.0cm]{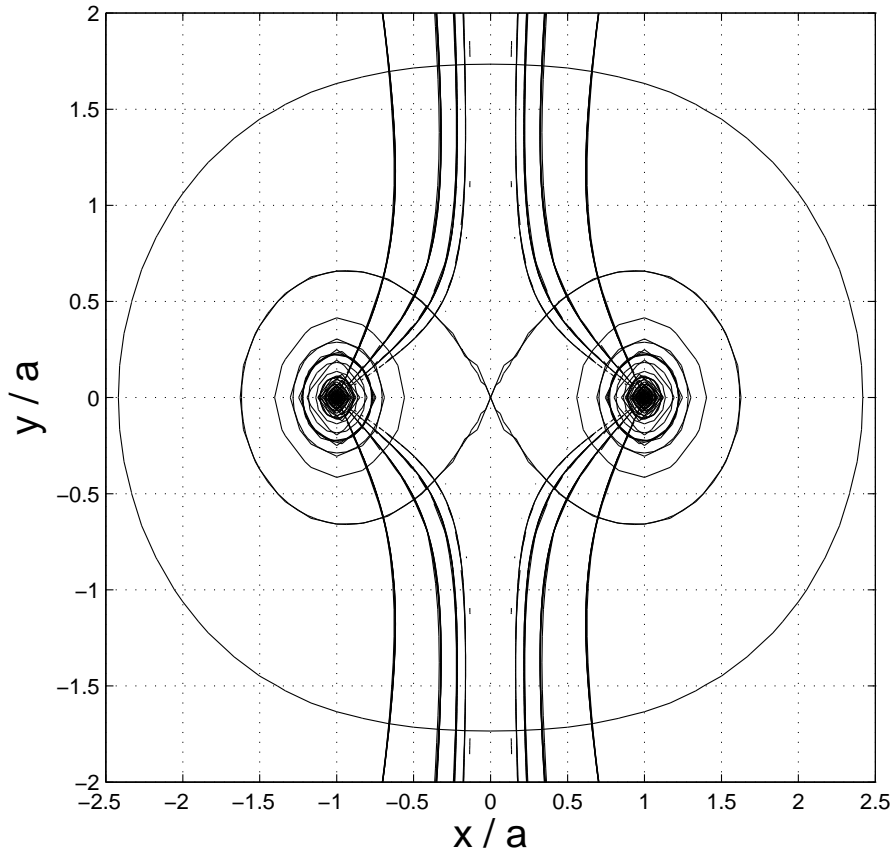}
\caption{The adapted coordinate system in the $x-y$ plane. Rotation about the $x$-axis yields the full 3D system. The closed contours are the equal-$\chi$ surfaces. The ``figure eight" curve is the critical surface across which the connectedness of the $\chi$ coordinate undergoes a topology change. The open curves are the equal-$\Theta$ lines.}
\label{fig2}
\end{figure}

In searching for the adapted coordinates we follow a similar path: First, define the coordinate $\chi$ so that equal-$\chi$ surfaces are also equipotential surfaces of the static 3D Coulomb problem. Because of the problem's linearity, we may superpose solutions of two single-charge problems, for which the potentials (of unit charges) are $1/L$ and $1/R$, respectively, i.e., the total potential is $1/L+1/R$. 
In fact, {\em any} monotonic function of $1/L+1/R$ is a good choice for this coordinate, i.e., $\chi=\chi(1/L+1/R)$. Specifically, we choose $\chi$ so that at great distances ($L,R\gg a$), the coordinate $\chi$ behaves like the radial coordinate $r$, because in this limit the potential is almost that of a point particle (of charge $2q$). That is, we choose 
\begin{equation}
\frac{1}{\chi}:=\frac{1}{2}\,\left(\frac{1}{R}+\frac{1}{L}\right)
\end{equation}
or 
\begin{equation}
\chi=2\,\frac{LR}{L+R}\, .
\end{equation}
Notice that the coordinate $\chi$ is non-trivial: it undergoes a topology change at the critical surface $\chi=a$. For $\chi <a$ equal-$\chi$ surfaces are doubly connected, and for $\chi >a$ they are singly connected. Figure \ref{fig2} shows the equal-$\chi$ surfaces. 

To find the coordinate $\Theta$ we require that it is everywhere orthogonal to equal-$\chi$ surfaces (which are also the equipotential surfaces). Namely, equal-$\Theta$ lines will be the electric field lines. The
condition that two curves are orthogonal is that their gradients are. The reason why we need to require the condition of the gradients is that inner product is a map from two vectors to a scalar. Orthogonality of two vectors is defined by the requirement that their inner product vanishes. The way to create a vector from a scalar is by calculating its gradient. Specifically, $\Theta$ satisfies
\begin{equation}\label{pde}
\,\nabla\chi\cdot\,\nabla\Theta=0\, ,
\end{equation}
which in Cartesian coordinates is
\begin{equation}\label{pde-car}
\frac{\,\partial\chi}{\,\partial x}\frac{\,\partial\Theta}{\,\partial x}+\frac{\,\partial\chi}{\,\partial y}\frac{\,\partial\Theta}{\,\partial y}+\frac{\,\partial\chi}{\,\partial z}\frac{\,\partial\Theta}{\,\partial z}=0\, .
\end{equation}

The most straighforward way to try and solve the partial differential equation for $\Theta$ (recall that $\chi$ is known; the unknown in Eq.~(\ref{pde}) is $\Theta$) is to try and solve Eq.~(\ref{pde-car}) in cartesian coordinates. This turns out to be a highly non-trivial thing to do.  In Appendix B we discuss this equation. The easier way to solve Eq.~(\ref{pde}) is by realizing that it is a scalar equation, so that we can express the terms on either hand side in whatever coordinates we find convenient, and the equation will retain its form. Equation (\ref{pde}) becomes particularly simple when expressed in 3D--TDC. The reason why this could be expected is that space is symmetrical under the exchange of the two centers. It is therefore reasonable to expect that geometrical expressions would be simpler when expressed in terms of the distances to the two centers (and not other points such as the origin of the cartesian grid), the only ``natural" quantities that the metric structure of space depends on. 

We can express Eq.~(\ref{pde}) in 3D--TDC in two way: first, we can transform Eq.~(\ref{pde-car}) or we can use the metric (\ref{lr-metric}) to write $g^{ij}\chi_{,i}\Theta_{,j}=0$, a comma denoting partial derivative. Either way, we find that 
Eq.~(\ref{pde}) transforms to
\begin{equation}
\left[ \frac{L}{2}(L^2+R^2-4a^2)  +R^3 \right] L\,\frac{\,\partial\Theta}{\,\partial L}+\left[ \frac{R}{2}(L^2+R^2-4a^2)  +L^3 \right] R\,\frac{\,\partial\Theta}{\,\partial R}=0
\end{equation}
whose solution is
\begin{equation}\label{theta1}
\Theta(L,R) = \Theta\left\{   \frac{R-L}{RL} \, \left[(R+L)^2-4a^2\right]  \right\}\, .
\end{equation}
This solution is easy to verify, or even derive using software such as Maple or Mathematica. Notice that the solution is {\em any} function of a certain combination of $L,R$, with $a$ acting as a parameter (the only length scale in this problem). Indeed, any function of this argument is everywhere orthogonal to equal-$\chi$ surfaces. To make our  solution useful, we next need to choose judiciously which function. Specifically, we make our choice by requiring that the asymptotic properties of the solution for $\Theta$ will coincide with the regular polar coordinate $\theta$. This way, our coordinate system will asymptotically approach regular spherical coordinates as it should, as at great distances the separation of the two centers becomes negligible.

At very great distances, 
$$\frac{R-L}{RL} \, \left[(R+L)^2-4a^2\right]\to -8a\frac{x}{y}+O(y^{-3})\, .$$
We next require that at that limit a coordinate $\tilde \Theta$ coincides with the regular polar coordinate $\theta:=\tan^{-1}(y/x)$, which motivates us to choose
\begin{equation}
\tilde \Theta(L,R) = \tan^{-1}\left[  \frac{LR}{L-R} \, \frac{8a}{(R+L)^2-4a^2}  \right]\, .
\end{equation}
This guarantees that at very great distances $\tilde\Theta$ behaves similarly to the regular polar coordinate $\theta$. However, the coordinate $\tilde \Theta$ is still not quite what we need, because it is discontinuous: Consider a semicircle at constant large distance from the center of the cartesian coordinate system (the center is half way between the two charges), starting on the positive $x$-axis and going through the upper half-plane to the negative $x$-axis. The value of $\tilde\Theta$ will vary from $\pi/4$ to $\pi/2$, jump discontinuously to $-\pi/2$ crossing the $y$-axis, and then change to $-\pi/4$. Define then
\begin{equation}\label{theta}
\Theta(L,R):=\frac{\pi}{2}-\pi\,{\rm sgn}(L-R)+2\tilde\Theta(L,R)\, .
\end{equation}
The coordinate $\Theta$ is continuous in the upper half-plane, and it is everywhere orthogonal to $\chi$. It's range is from $0$ to $\pi$. Notice that ${\rm sgn}(L-R)\equiv{\rm sgn}(x)\equiv{\rm sgn}(\pi/2-\Theta)$, and that we only need to define $\Theta$ in the upper half-plane, because we rotate the coordinates  about the $x$-axis. 

We now have the coordinate system $(\chi,\Theta,\Phi)$, which is the Two-Center Bi-Spherical coordinates system (TCBS). The metric in these coordinates is given by
\begin{eqnarray}\label{TCBS}
\,d\sigma^2 &=& \frac{1}{4}\,\frac{(L+R)^4}{(L+R)^4-3LR(L+R)^2-4LRa^2}\,d\chi^2 \nonumber \\
&+&\frac{1}{16\cdot 256}\, \frac{[16(L-R)^2a^4-8(L^4+R^4-10L^2R^2)a^2+(L-R)^2(L+R)^4]^2}{a^4\,Q^2\,[(L+R)^4-3LR(L+R)^2-4LRa^2]}\,d\Theta^2 \nonumber \\
&+& Q^2\,d\Phi^2\, ,
\end{eqnarray}
where $L,R$ are implicit functions of $\chi,\Theta$. The metric (\ref{TCBS}) is manifestly diagonal. 
Given $\chi,\Theta$, one needs to solve the following quintic to find $R(\chi,\Theta)$ explicitly:
\begin{equation}
R^5-\chi\,R^4+4a(h^{-1}\chi-a)\,R^3-4a\chi (h^{-1}\chi-2a)\,R^2+a\chi^2(h^{-1}\chi-5a)\,R+a^2\chi^3=0\,
\end{equation}
where
$$h :=\,\tan {\tilde {\Theta}}=\,\tan \frac{1}{2}\left[  \Theta-\frac{\pi}{2}  +\pi\,{\rm sgn}\left(\frac{\pi}{2}-\Theta\right)  \right]\, .$$ 
As is well known from Abel's Impossibility Theorem \cite{Abel}, it is impossible to solve a general quintic equation in terms of radicals. 
However, solutions in terms of hypergeometric functions \cite{Klein} or Jacobi Theta functions \cite{Hermite} are always possible. Numerical solutions are of course easy to find (e.g., using the Newton--Raphson method). Notice, that the Fundamental Theorem of Algebra guarantees a real solution.  
It is immediately clear that the coordinate system is {\em singular} at the two centers and at the origin of the cartesian coordinate system. For example, $Q^2$ vanishes at all three singular points, so that $g_{\Theta\Theta}$ diverges, and $g_{\Phi\Phi}$ vanishes.  As space is 3D flat Euclidean space, we know that this singularity is a coordinate singularity, and not a genuine geometrical one. Notably, the Jacobian of the transformation from cartesian to 3D--TCBS coordinates is regular. 


Given the solution for $R(\chi,\Theta)$ we can readily find $L(\chi,\Theta)$:
\begin{equation}
L=\frac{\chi\,R}{2R-\chi}\, .
\end{equation}


\section{The potential problem in the adapted coordinates}

To express the potential problem in the newly found TCBS coordinates we first express the Laplacian in these coordinates. In any coordinate system the Laplacian of a scalar field $\Psi$ is given by
\begin{eqnarray}\label{laplacian}
\,\nabla^2\Psi &=& (\,\nabla\chi\cdot\,\nabla\chi)\; \Psi_{,\chi\chi} +
 (\,\nabla\Theta\cdot\,\nabla\Theta)\; \Psi_{,\Theta\Theta}+
 ( \,\nabla\Phi\cdot\,\nabla\Phi)\; \Psi_{,\Phi\Phi} \nonumber \\
  &+&  2(\,\nabla\chi\cdot\,\nabla\Theta)\; \Psi_{,\chi\Theta}+
  2 (\,\nabla\chi\cdot\,\nabla\Phi)\; \Psi_{,\chi\Phi}+
   2 (\,\nabla\Theta\cdot\,\nabla\Phi)\; \Psi_{,\Theta\Phi} \nonumber \\
   &+& (\,\nabla^2\chi) \;\Psi_{,\chi}+ (\,\nabla^2\Theta) \;\Psi_{,\Theta}+  (\,\nabla^2\Phi) \;\Psi_{,\Phi}\, .
\end{eqnarray}
The coefficients in this expression can readily be calculated explicitly from the above expressions. Specifically, we find
\begin{equation}\label{lap-chi}
\,\nabla^2\chi=4\,\frac{L^4+R^4+LR(L^2+R^2-4a^2)}{LR(L+R)^3}
\end{equation}
\begin{equation}
\,\nabla^2\Theta = \frac{128\,a^3\,(L-R)\, G}{[16(L+R)^2a^4+8(L^4-10L^2R^2+R^4)a^2-(L-R)^2(L+R)^4]^2}
\end{equation}
where
\begin{eqnarray}
G &=& 64a^6-16(5L^2+2LR+5R^2)a^4+4(7L^4+28L^3R+26L^2R^2+28LR^3+7R^4)a^2\nonumber \\
&-& (L-R)^2(3L^4+20L^3R-14L^2R^2+20LR^3+3R^4)\, , 
\end{eqnarray}
\begin{equation}
\,\nabla^2\Phi=0
\end{equation}
\begin{equation}
\,\nabla\chi\cdot\,\nabla\chi=g^{\chi\chi}
\end{equation}
\begin{equation}
\,\nabla\Theta\cdot\,\nabla\Theta=g^{\Theta\Theta}
\end{equation}
\begin{equation}
\,\nabla\Phi\cdot\,\nabla\Phi=g^{\Phi\Phi}
\end{equation}
\begin{equation}
\,\nabla\chi\cdot\,\nabla\Theta=0
\end{equation}
\begin{equation}
\,\nabla\chi\cdot\,\nabla\Phi=0
\end{equation}
\begin{equation}
\,\nabla\Theta\cdot\,\nabla\Phi=0\, ,
\end{equation}
the last three relations resulting from the TCBS coordinates being orthogonal. Notice that $\Phi$ is harmonic.  

We are now in a position to show that in the 3D--TCBS the solution for the potential problem of two equal like charges is given by a monopole term only. Specifically, we show that 
\begin{equation}\label{lap_1chi}
\,\nabla^2 \left(\frac{1}{\chi}\right)=0\;\;\;\;\;\;\;\;\;\;\;\;(\chi\ne 0)\, .
\end{equation}
This relation can be directly verified by substitution in Eq.~(\ref{laplacian}): Clearly all the derivatives with respect to either $\Theta$ or $\Phi$ vanish, so that Eq.~(\ref{laplacian}) reduces to 
\begin{eqnarray}
\,\nabla^2  \left(\frac{1}{\chi}\right) &=& g^{\chi\chi} \left(\frac{1}{\chi}\right)_{,\chi\chi}+\,\nabla^2\chi \left(\frac{1}{\chi}\right)_{,\chi}\nonumber \\
&=& \frac{2}{\chi^3}g^{\chi\chi}-\frac{\,\nabla^2\chi}{\chi^2}=0
\end{eqnarray}
after direct substitution of $g^{\chi\chi}$ from Eq.~(\ref{TCBS}) and using Eq.(\ref{lap-chi}). Equation 
(\ref{lap_1chi}) is analogous to $\,\nabla^2 (1/r)=0$ (except at $r=0$) in regular spherical coordinates.  
At $\chi=0$ ($x=\pm a,y=z=0$) the Laplacian of $\chi$ no longer vanishes, so that the global problem is described by Poisson's equation, $\,\nabla^2\Psi=-8\pi q\,J^{-1}\,\delta(\chi)$, where $J$ is the Jacobian determinant for the transformation from cartesian to 3D--TCBS coordinates. Comparing Poisson's equation with Eq.~(\ref{lap_1chi}), we find the solution for the potential problem to be 
\begin{equation}
\Psi=\frac{2q}{\chi}\, ,
\end{equation}
which is the desired solution.


\section{Conclusions}

We showed, for the specific example of two equal like electric charges, how to find a coordinate system in which the electric potential is described by one coordinate, and for which it is given by a monopole term only. The derivation involved a large number of topics that are covered in mathematical methods courses, such as non-diagonal coordinate systems, coordinate singularities, quintic equations, multipole expansions, coordinate transformations, and potential theory. As such, it may serve as an instructive problem for such courses. In particular, it may be used to demonstrate the deep meaning of a multipole expansion, and its dependence on the choice of coordinates. Specifically, a multipole expansion can be made simple when the coordinates used are adapted to the potential problem. The reason why an infinite number of (even) multipoles are needed to describe the potential of two equal like charges using regular spherical coordinates (\ref{multipoles}) is that the equal coordinate surfaces are very different from the equipotential surfaces. By choosing the two surfaces to coincide, we are able to eliminate all the higher moltipoles from the potential, and solve the potential problem using only the monopole term. The price to pay is in a form of more involved mathematics, which is however still in the range of the usual preparation of the usual Physics education programs. 
 
\section*{Acknowledgments}

The author wishes to thank Richard Price and Anthony Hester for discussions, and Ross Cortez for checking some of the calculations. This work was supported in part by a minigrant from the UAH Office of the Vice President for Research. 

\begin{appendix}

\section{``Derivation" of spherical coordinates}

In this Appendix we apply our method to a point charge, and derive the 
adapted coordinates, which are just the regular spherical coordinates. This derivation may serve as 
a pedagogic illustration of the method, applied to a trivial situation for which the solution is well known. 

As discussed above, for the case of a single point particle, the radial coordinate $r$ describes equipotential surfaces. We therefore choose the radial coordinate $r$ as one of the 
adapted coordinates, or $\chi_{\rm s} :=r$.  To find the second coordinate, we require that it is orthogonal to $r$. Therefore, it satisfies Eq. ~(11), i.e., $\,\nabla\chi_{\rm s}\cdot\,\nabla\Theta_{\rm s}=0\, ,$
or, in cartesian coordinates in the $x$--$y$ plane,
$({\,\partial r}/{\,\partial x})({\,\partial\Theta_{\rm s}}/{\,\partial x})+({\,\partial r}{\,\partial y})/({\,\partial\Theta_{\rm s}}{\,\partial y})=0$.
Recalling that $\,\partial r /\,\partial x=x/r$ and $\,\partial r /\,\partial y=y/r$, this equation becomes
$x\,({\,\partial\Theta_{\rm s}}/{\,\partial x})+y\,({\,\partial\Theta_{\rm s}}/{\,\partial y})=0$ (except for the origin which is a coordinate singularity), whose solution is $\Theta_{\rm s}=\Theta_{\rm s}(y/x)$. Next, we require that at large distances this coordinate coincides with the regular azimuthal coordinate 
$\theta:=\tan^{-1}(y/x)$, so that we find that everywhere $\Theta_{\rm s}:=\theta$.

Rotating the coordinates by an angle $\phi$ about the $x$-axis yields 
$x'=x=r\,\cos\theta$, $y'=y\,\cos\phi=r\,\sin\theta\,\cos\phi$, and $z'=y\,\sin\phi=r\,\sin\theta\,\sin\phi$. The regular spherical coordinates are obtained by renaming the cartesian axes $x''=y'$, $y''=z'$, and $z''=x'$: 
$x=r\,\sin\theta\,\cos\phi$, $y=r\,\sin\theta\,\sin\phi$, and $z=r\,\cos\theta$, after dropping the primes for 
conventionality.

\section{Trying to solve Eq.~(\ref{pde-car}) directly}

Equation (\ref{pde-car}) can be readily written explicitly. In the $x-y$ plane, the unknown is $\Theta (x,y)$, so that Eq.~(\ref{pde-car}) becomes
\begin{eqnarray*}
\left[ (x-a)\left( x^2+2ax+a^2+y^2\right)^{3/2}+ (x+a)\left(x^2-2ax+a^2+y^2\right)^{3/2}\right]\frac{\,\partial\Theta}{\,\partial x}\\
+ y\left[ \left( x^2+2ax+a^2+y^2\right)^{3/2}+\left(x^2-2ax+a^2+y^2\right)^{3/2}\right]
\frac{\,\partial\Theta}{\,\partial y}=0\, .
\end{eqnarray*}
This partial differential equation is non-trivial to solve symbolically. Standard techniques, e.g., separation of variables, prove to be ineffective. Even PDE solvers such as Maple and Mathematica, using up to 8 gigabytes RAM for a couple of weeks on a PowerMac G5 were unsuccessful in solving this equation. Indeed, having found the solution (\ref{theta1}) using the 3D--TDC it is evident why straightforward attempts to solve this equation failed. It should be noted, however, that numerical solutions for this equation can readily be found, if appropriate boundary conditions are specified.

\end{appendix}



\begin{thebibliography}{99}

\bibitem{Abel} N.H.~Abel, ``Beweis der Unm\"{o}glichkeit, algebraische Gleichungen von h\"{o}heren Graden als dem vierten allgemein aufzul\"{o}sen," J. reine angew. Math. {\bf 1}, 65 (1826).

\bibitem{Arfken} G.B Arfken and H.J. Weber, {\em Mathematical Methods for Physicists}, 6$^{\rm th}$ ed. (Academic Press, San Diego, CA, 2005). 

\bibitem{Griffiths} D.J. Griffiths, {\em Introduction to Electrodynamics}, 3$^{\rm rd}$ ed. (Prentice Hall, Upper Saddle River, NJ, 1999). 

\bibitem{Franklin} J. Franklin, {\em Classical Electromagnetism} (Pearson -- Addison Wesley, San Francisco, CA, 2005). 

\bibitem{Hermite} C. Hermite, ``Sulla risoluzione delle equazioni del quinto grado," Annali di math. pura ed appl. {\bf 1}, 256--259 (1858).

\bibitem{Jackson} J.D. Jackson, {\em Classical Electrodynamics}, 3$^{\rm rd}$ ed. (John Wiley \& Sons, New York, NY, 1999).



\bibitem{Klein} F. Klein, {\em Lectures on the Icosahedron and the Solution of Equations of the Fifth Degree}   (Dover Publications, New York, NY, 1956).




\bibitem{Mathews} J. Mathews and R.L. Walker, {\em Mathematical Methods of Physics}, 2$^{\rm nd}$ ed. (Addison--Wesley, Redwood City, CA, 1970).



\bibitem{Morse} P.M. Morse and H. Feshbach, {\em Methods of Theoretical Physics}  (McGraw--Hill, New York, NY, 1953).

\bibitem{Panofsky} W.K.H. Panofsky and M. Phillips, {\em Classical Electricity and Magnetism}, 2$^{\rm nd}$ ed. (Addison-Wesley, Reading, MA, 1962). 

\bibitem{Pollack} G.L. Pollack and D.R. Stump, {\em Electromagnetism} (Addison Wesley, San Francisco, CA, 2002). 

\bibitem{Reitz} J.R. Reitz, F.J. Milford, and R.W. Christy, {\em Foundations of Electromagnetic Theory}, 4$^{\rm th}$ ed. (Addison Wesley, Reading, MA, 1992).

\end{thebibliography}
\end{document}